\documentclass[useAMS,usenatbib]{mn2e}

\usepackage{epsfig}
\usepackage{longtable}
\usepackage{times} 
\newcommand{\sw}{$Swift$}

\def \igr {\mbox{IGR~J17544$-$2619}}
\def \src {\mbox{XTE~J1739$-$302}}

\def \inte {{\em INTEGRAL}}
\def \rxte {{\em RXTE}}
\def \sax {{\em BeppoSAX}}
\def \suzaku {{\em Suzaku}}
\def \sw {{\em Swift}}

\def \ne {n_{\rm e}}

\def \hcm {\hbox {\ifmmode $ atom cm$^{-2}\else atom cm$^{-2}$\fi}}

\def \chiq {$\chi^{2}$}

\def \sax{{\it BeppoSAX}}
\def\ktbb{kT_{\rm bb}}
\def\ktw{kT_{\rm w}}
\def\kte{kT_{\rm e}}
\def\scw{Schwarzschild}

\def \wabs {\textsc{wabs}} 
\def \edge {\textsc{edge}}
\def \cpl {\textsc{cutoffpl}}
\def \hcp {\textsc{highecut}}
\def \bb {\textsc{bbodyrad}} 
\def \comptt {\textsc{comptt}}
\def \compmag {\textsc{compmag}}

\def\b0{\beta_{\rm 0}}

\def \ATel {Astron.\ Tel.}
\def \apj {ApJ}
\def \apjl {ApJL}
\def \aap {A\&A}

\def \mnras {MNRAS}

\newcommand{\pder}[2]{\frac{\partial #1}{\partial #2}}
\newcommand{\tder}[2]{\frac{d #1}{d #2}}
\newcommand{\be}{\begin{equation}}
\newcommand{\ee}{\end{equation}}
\title[SFXTs with {\it Swift}: outbursts of prototypes]{\emph{Swift} observations of two supergiant fast \mbox{X-ray} transient prototypes in outburst}

\author[R.\ Farinelli et al.]{R. Farinelli$^{1,2}$, P.\ Romano$^{2}$, V.\ Mangano$^{2}$, C.\ Ceccobello$^{1}$,  L.\ Ducci$^{3}$, 
\newauthor S.\ Vercellone$^{2}$, P.\ Esposito$^{4}$, J.A.~Kennea$^{5}$, D.N.~Burrows$^{5}$ \\ 
$^{1}$Dipartimento di Fisica, Universit\`a di Ferrara, via Saragat 1, 44122, Ferrara, Italy\\
$^{2}$INAF, Istituto di Astrofisica Spaziale e Fisica Cosmica,
        via U.\ La Malfa 153, 90146 Palermo, Italy\\
$^{3}$  Institut f\"ur Astronomie und Astrophysik,
         Universit\"at T\"ubingen, Sand 1, 72076 T\"ubingen, Germany \\
$^{4}$INAF, Osservatorio Astronomico di Cagliari, localit\`a Poggio dei Pini, strada 54, 09012 Capoterra, Italy\\
$^{5}$Department of Astronomy and Astrophysics, Pennsylvania State 
             University, University Park, PA 16802, USA\\
}

\begin{document}

\date{Accepted 2012 May 30.  Received 2012 May 30; in original form 2012 April 17}

\pagerange{\pageref{firstpage}--\pageref{lastpage}} \pubyear{2012}

\maketitle

\label{firstpage}

\begin{abstract}
\noindent
We report on the results from observations of the most recent outbursts of \src\ and \igr, 
which are considered to be the prototypes of the supergiant fast \mbox{X-ray} transient (SFXT) class.
They triggered the {\it Swift}/BAT on 2011 February 22 and March 24, respectively,  
and each time a prompt {\it Swift} slew allowed us to obtain the rich broad-band data we present. 
The XRT light curves show the descending portion of very bright flares that reached 
luminosities of $\sim 2\times10^{36}$ and $\sim 5\times10^{36}$ erg s$^{-1}$, respectively. 
The broad-band spectra, when fit with the usual phenomenological models adopted for accreting 
neutron stars, yield values of both high energy cut-off and e-folding energy 
consistent with those obtained from previously reported outbursts from these sources.
In the context of more physical models, the spectra of both sources can be
well fitted either with a two-blackbody model, or with a single
unsaturated Comptonization model.
In the latter case, the model can be either a classical static Comptonization model,
such as \comptt, or the recently developed \compmag\ model, which includes thermal
and bulk Comptonization for cylindrical accretion onto a magnetized neutron star.
We discuss the possible accretion scenarios derived by the different models,
and we also emphasize the fact that the electron density derived from
the Comptonization models, in the regions where  the \mbox{X-ray} spectrum presumably forms,
is lower than that estimated using the continuity equation at the magnetospheric radius 
and the source \mbox{X-ray} luminosity, and we give some possible explanations.

\end{abstract}

\begin{keywords}
\mbox{X-rays}: binaries -- \mbox{X-rays}: individual: \src, \igr. 

\noindent
Facility: {\it Swift}

\end{keywords}

%%%%%%%%%%%%%%%%%%%%%%%%%%%%%%%%%%%%%%%%%%%%%%%%

%%%%%%%%%%%%%%%%%%%%%%%%%%%%%%%%%%%%%%%%%%%%%%%%%%%%%%%%   TABLE 1
\begin{table*}
 \begin{center}
 \caption{Log of the observations of \src\ and \igr\ with {\it Swift}.\label{tab:observations} }
 \begin{tabular}{lllllrr}
 %\hline
 \hline
 \noalign{\smallskip}
Source &  Sequence & Instrument & Start time (UT) & End time (UT) &  Exposure & Time Since Trigger\\ %$^{\mathrm{a}}$
  & & /Mode & (yyyy-mm-dd hh:mm:ss) & (yyyy-mm-dd hh:mm:ss) & (s) & (s) \\
  \noalign{\smallskip}
 \hline
 \noalign{\smallskip}
\src    &00446475000	&BAT/evt	&2011-02-22 07:19:44	&2011-02-22 07:36:46	&1022 & -119	\\
                   &00446475000	&XRT/WT	&2011-02-22 07:24:10	&2011-02-22 09:04:37	&1271 & 147	\\
                   &00446475000	&XRT/PC	&2011-02-22 09:04:39	&2011-02-22 09:13:55	&554  & 6176	\\
\igr  &00449907000	&BAT/evt	&2011-03-24 01:53:04	&2011-03-24 02:13:06	&1202 & -239	\\
                   &00449907000	&XRT/WT	&2011-03-24 01:59:15	&2011-03-24 03:18:22	&657  & 133     \\
                   &00449907000	&XRT/PC	&2011-03-24 02:10:07	&2011-03-24 04:01:03	&3360 & 784     \\
                   &00035056150	&XRT/PC	&2011-03-27 19:54:23	&2011-03-27 20:11:56	&1046 & 323840	\\
                   &00035056151	&XRT/PC	&2011-03-28 20:10:36	&2011-03-28 20:21:58	&662  & 411213	\\
                   &00035056152	&XRT/PC	&2011-03-29 20:15:13	&2011-03-29 20:23:57	&504  & 497890	\\
  \noalign{\smallskip}
  \hline
  \end{tabular}
  \end{center}
\end{table*} %%%%%%%%%%%%%%%%%%%%%%%%%%%%%%%%%%%%%%%%

\section{Introduction}
\label{intro}

Supergiant fast X--ray transients (SFXTs) are a class of High Mass
X--ray Binaries (HMXBs) associated with OB supergiant stars.
In the X--rays they display outbursts significantly shorter 
than those of typical Be/X--ray binaries characterized by bright flares 
with peak luminosities of 10$^{36}$--10$^{37}$~erg~s$^{-1}$ which  
last a few hours \citep[as observed by \inte; ][]{Sguera2005,Negueruela2006:ESASP604}.
As their quiescence is characterized by a luminosity of $\sim 10^{32}$~erg~s$^{-1}$ 
\citep[e.g.\ ][]{zand2005,Bozzo2010:quiesc1739n08408}, 
their dynamic range is of 3--5 orders of magnitude. 
While in outburst, their hard X--ray spectra resemble those of HMXBs 
hosting an accreting neutron star (NS), with hard power laws below 10\,keV  
with high energy cut-offs at $\sim 15$--30~keV.  
So, even if pulse periods have only been measured for a few SFXTs,
it is tempting to assume that all SFXTs might host a NS. 

The physical context originating the outbursts has been claimed to be related 
either to the properties of  the wind from the supergiant companion 
\citep{zand2005,Walter2007,Negueruela2008,Sidoli2007} or to the 
presence of a centrifugal or magnetic barrier \citep[][]{Grebenev2007,Bozzo2008}. 

\src\ was discovered by \rxte\ in August 1997 \citep{Smith1998:17391-3021}, 
when it reached a peak flux of 3.6$\times$10$^{-9}$~erg cm$^{-2}$~s$^{-1}$ (2--25 keV).
It has a long history of flaring recorded by \inte\ \citep{Sguera2006,Walter2007,Blay2008}  
and by \sw\ \citep{Sidoli2009:sfxts_paperIII,Sidoli2009:sfxts_paperIV,Romano2011:sfxts_paperVI,Romano2011:atel3182}. 
Recently, \citet{Drave2010:17391_3021_period} reported the discovery of a $51.47\pm0.02$\,d 
orbital period based on $\sim 12.4$\,Ms of \inte\ 
data\footnote{We note, however, that \citet[][]{Romano2009:sfxts_paperV} derived 
a marginal evidence, based on XRT data, of signal at $P_{\rm orb}=12.8658\pm0.0073$\,d,
or $1/4$ the period reported by \citet{Drave2010:17391_3021_period}.}. 
The optical counterpart is an O8I ab star at 2.7\,kpc \citep{Negueruela2006,Rahoui2008}.

\igr\ was first detected by \inte\ in 2003 \citep{Sunyaev2003}, 
when the source reached a flux of 160~mCrab (18--25~keV). 
Several more flares, lasting up to 10 hours, were detected by \inte\ in  the following years 
\citep{Grebenev2003:17544-2619,Grebenev2004:17544-2619,Sguera2006,Walter2007,Kuulkers2007} 
with fluxes up to 400~mCrab (20--40 keV); some were also found in archival \sax\ observations 
\citep{zand2004:17544bepposax}. 
Subsequent flares were observed by \sw\  
\citep{Krimm2007:ATel1265,Sidoli2009:sfxts_paperIII,Sidoli2009:sfxts_paperIV,Romano2011:sfxts_paperVI,Romano2011:sfxts_paperVII,Romano2011:atel3235}, 
and \suzaku\ \citep{Rampy2009:suzaku17544}. 
\citet{Clark2009:17544-2619period} reported the discovery of a $4.926\pm0.001$\,d 
orbital period based on $\sim 4.5$ years of \inte\ data. 
Recently, \citet[][]{Drave2012:17544_2619_pulsation} detected 
a transient $71.49\pm0.02$\,s signal
in RXTE observations from the region of \igr, which, if
interpreted as the spin period of the NS in the system, 
places the source in the locus of the Corbet diagram \citep[][]{Corbet1986} 
where classical wind-fed supergiant X--ray binaries can be found.    
The optical counterpart is an O9Ib star at 3.6\,kpc \citep{Pellizza2006,Rahoui2008}.

\src\  and \igr\ are considered the prototypes of the SFXT class, and
were extensively studied with \sw. 
In particular, in addition to the Burst Alert Telescope \citep[BAT, ][]{Barthelmy2005:BATmn}
outburst detections and intensive  \mbox{X-ray} Telescope  \citep[XRT, ][]{Burrows2005:XRTmn} follow-up, \sw\ has been 
studying their long-term properties \citep[][and references therein]{Romano2011:sfxts_paperVI}. 
In this paper we examine the most recent outbursts of these two sources, which triggered the 
BAT in 2011. 
In particular, we apply, for the first time to SFXTs, 
the new {\sc COMPMAG} model by \citet[][hereafter F12]{Farinelli2012:compmag}.

\begin{figure}%%%%%%%%%%%%%%%%%%%%%%%%%%%%%%%%%%%%%%%%%%%%%%%%%%%%%   FIGURE 1
\begin{center}
\vspace{-2.9truecm}
\hspace{-0.4cm}
\centerline{\includegraphics[width=9.7cm,angle=0]{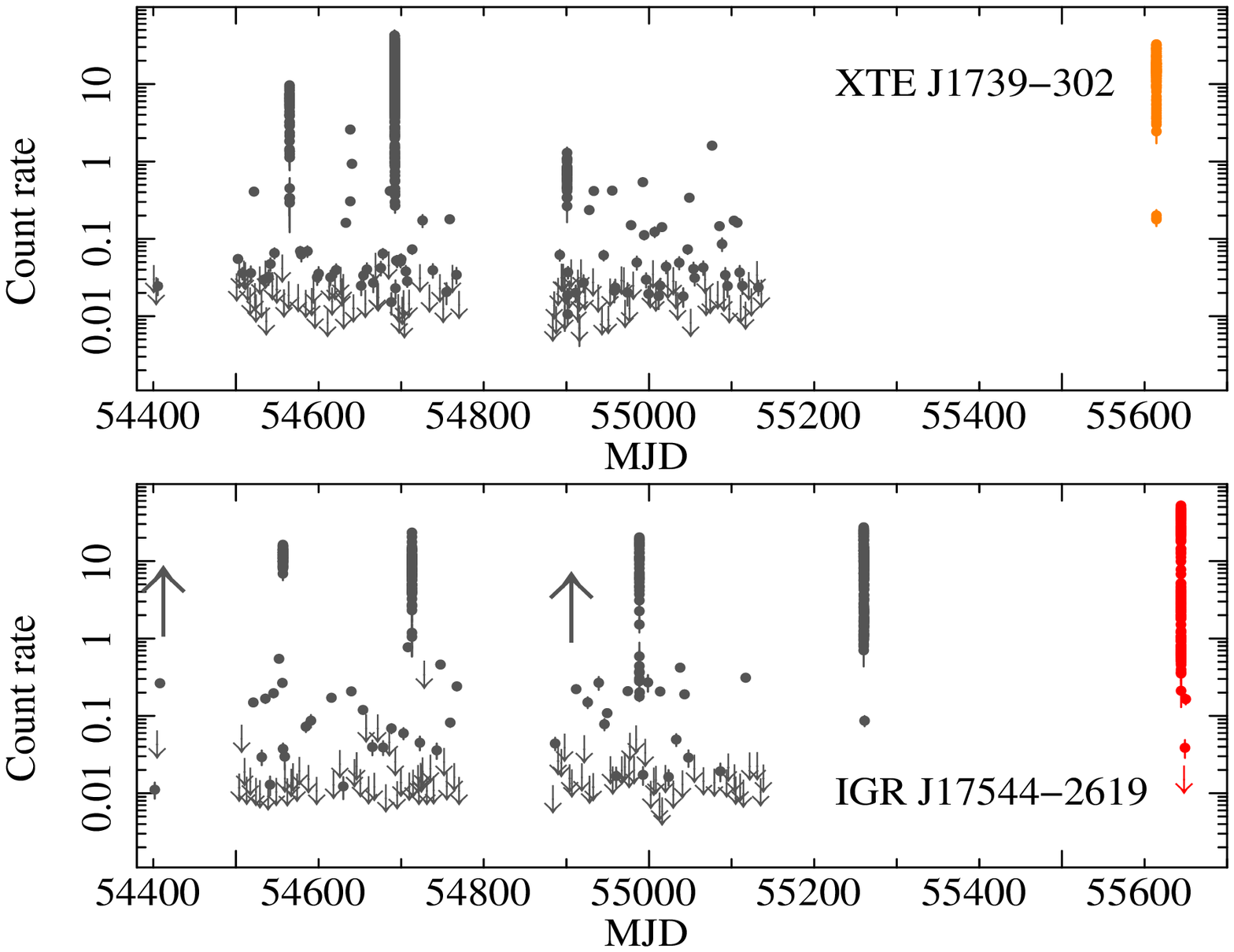}}
\vspace{-4.0truecm}
\caption[XRT light curves]{\sw/XRT (0.2--10\,keV) long-term light curves of \src\ and \igr.   
		The downward-pointing arrows are 3$\sigma$ upper limits. The upward-pointing arrows 
                mark flares that triggered the BAT Transient Monitor on MJD 54414 and 54906.  
                Data up to MJD $\sim 55261$ (grey) were published in 
             \citet[][]{Romano2009:sfxts_paperV,Romano2011:sfxts_paperVI,Romano2011:sfxts_paperVII}. }
		\label{fig:xrtlcvs} 
\end{center}
\end{figure}%%%%%%%%%%%%%%%%%%%%%%%%%%%%%%%%%%%%%%%%%%%%%%%%%%%%% 

\begin{figure}%%%%%%%%%%%%%%%%%%%%%%%%%%%%%%%%%%%%%%%%%%%%%%%%%%%%%   FIGURE 2
\begin{center}
\vspace{-2.9truecm}
\hspace{-0.4cm}
\centerline{\includegraphics[width=9.7cm,angle=0]{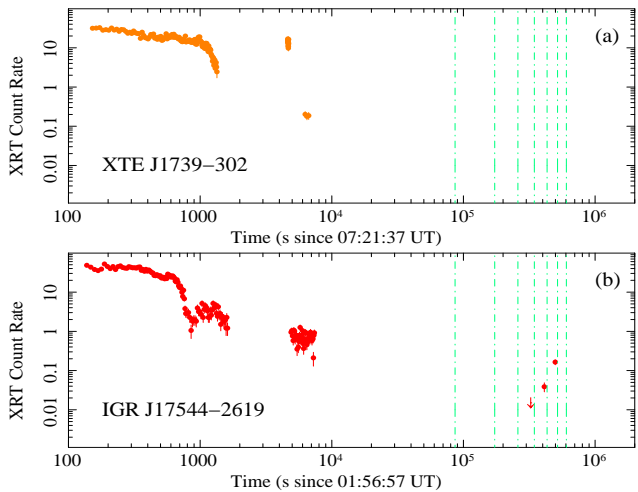}}
\vspace{-4.0truecm}
\caption[XRT light curves, zoom]{\sw/XRT (0.2--10\,keV) light curves of the
  2011 outbursts of \src\ and \igr, as 
  followed by {\it Swift}/XRT, referred to their respective BAT triggers (2011 February 22, 2011 March 24).
  Points denote detections, downward-pointing arrows 3$\sigma$ upper limits.  
  Where no data are plotted, no \sw\ data were collected.  
  Vertical dashed lines mark time intervals equal to 1 day, up to a week. }
\label{fig:xrtlcvs2} 
\end{center}
\end{figure}%%%%%%%%%%%%%%%%%%%%%%%%%%%%%%%%%%%%%%%%%%%%%%%%%%%%% 

	%%%%%%%%%%%%%%%%%%%%%%%%%%%%%%%%%%%%%%%%%%%%%%%%%%%%%%%%%
	\section{Observations  and Data Reduction} \label{dataredu}
	%%%%%%%%%%%%%%%%%%%%%%%%%%%%%%%%%%%%%%%%%%%%%%%%%%%%%%%%%

\src\ triggered the BAT on 2011 February 22 at 07:21:37 UT 
\citep[image trigger=446475, ][]{Romano2011:atel3182}.  \sw{} immediately slewed to the target, 
so that the narrow-field instruments (NFI) started observing about 141\,s after the trigger. 
After the initial automated target (AT) observation (two orbits 
for a total of $\sim1.8$\,ks net exposure spanning $\sim6.7$\,ks) 
no further NFI observations were performed. 

\igr\ \citep[][]{Romano2011:atel3235} triggered the BAT on 2011 March 24 at 01:56:57 UT 
\citep[image trigger=449907, ][]{Romano2011:atel3235}.  \sw{} immediately slewed to the target, 
so that the NFIs started observing about 126\,s after the trigger. 
The AT (sequence 00449907000) ran for two orbits, until $\sim7.4$\,ks after the trigger).  
Follow-up target of opportunity (ToO) observations for a total of 2.2\,ks were obtained  
(sequences 00035056150--152) after the source emerged from Moon constraint. 
The data cover the first 6\,d after the beginning of the outburst.  
Table~\ref{tab:observations} reports the log of the \sw\ observations used in this paper.

The BAT 
data were analysed using the standard BAT analysis software 
within \textsc{ftools} in the \textsc{heasoft} package (v.6.11). 
Mask-tagged BAT light curves were created in the standard energy bands,
and rebinned to fulfil at least one of the following conditions, 
achieving a signal-to-noise (S/N) of 5 or bin length of 10\,s. 
Spectra were extracted from the whole event lists and within  
time intervals strictly simultaneous with the XRT. 
Response matrices were generated with \textsc{batdrmgen} 
using the latest spectral redistribution matrices. 
Survey data products, in the form of Detector Plane
Histograms (DPH), are available, and were analysed with the
standard \textsc{batsurvey} software.

The XRT data were processed with standard procedures 
(\textsc{xrtpipeline} v0.12.6), filtering and screening criteria by using 
\textsc{ftools}. 
We considered both windowed-timing (WT) and photon-counting (PC) mode data,
and selected event grades 0--2 and 0--12, 
for WT and PC data, respectively (\citealt{Burrows2005:XRTmn}). 
When appropriate, we corrected for pile-up 
by determining the size of the affected core of the point spread function (PSF)  
by comparing the observed and nominal PSF \citep{vaughan2006:050315mn},
and excluding from the analysis all the events that fell within that
region. 
Background events were accumulated from source-free regions. 
For our timing analysis light curves were created for several values of 
signal-to-noise ratio (SNR) and number of counts per bin; 
all were corrected for PSF losses, vignetting and background. 
For our spectral analysis, we extracted events in the same regions as 
those adopted for the light curve creation; the data were rebinned with a minimum 
of 20--50 counts per energy bin, as appropriate, to allow $\chi^2$ 
fitting, with the exception of low-count-statistics spectra, in which Cash 
\citep{Cash1979} statistics and spectra binned to 1 count per bin were used, instead.  
Ancillary response files were generated with \textsc{xrtmkarf},
to account for different extraction regions, vignetting, and PSF corrections. 
We used the latest spectral redistribution matrices in CALDB (20110915). 
For the luminosity calculation we adopted a distance of 
2.7\,kpc and 3.6\,kpc for \src\ and \igr, respectively \citep{Rahoui2008}. 

All quoted uncertainties are given at 90\,\% confidence level for 
one interesting parameter unless otherwise stated. 
The spectral indexes are parametrised as $F_{\nu} \propto \nu^{-\alpha}$, 
where $F_{\nu}$ (erg cm$^{-2}$ s$^{-1}$ Hz$^{-1}$) is the 
flux density as a function of frequency $\nu$; 
we adopt $\Gamma = \alpha +1$ as the photon index, 
$N(E) \propto E^{-\Gamma}$ (ph cm$^{-2}$ s$^{-1}$ keV$^{-1}$).

%%%%%%%%%%%%%%%%%%%%%%%%%%%%%%%%%%%%%%%%%%%%%%%%%%   FIGURE 3
\begin{figure}
\begin{center}
\vspace{-0.5truecm}
\hspace{+1.0truecm}
\centerline{\includegraphics[angle=270,width=9.0cm]{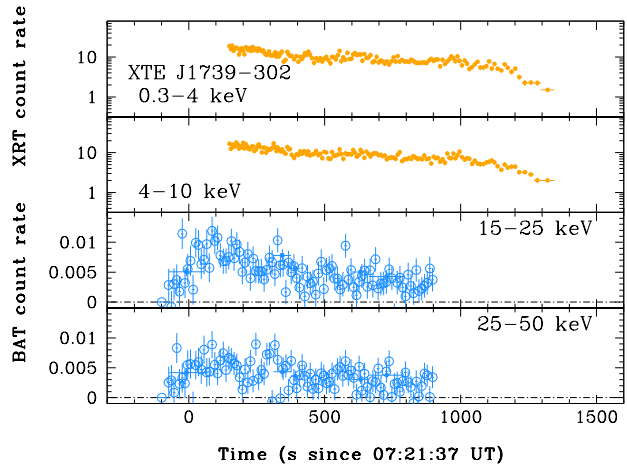}}
\end{center}
\caption{XRT and BAT light curves of the initial orbit of data of the 2011 February 22 outburst 
of \src\ in units of count s$^{-1}$ and count s$^{-1}$ detector$^{-1}$, respectively. 
The empty circles correspond to BAT in event mode,  
filled circles to BAT survey mode data. 
}
\label{fig:17391lcv_allbands}
\end{figure}
%%%%%%%%%%%%%%%%%%%%%%%%%%%%%%%%%%%%%%%%%%%%%%%%%%

 \begin{table}
\tabcolsep 3pt   
 \begin{center}
 \caption{Absorbed power-law spectral fits of XRT data of \src.}
 \label{tab:xrtband_17391}
 \begin{tabular}{lllllll}
% \hline
 \hline
 \noalign{\smallskip}
 Spectrum$^a$      &$N_{\rm H}^b$  &$\Gamma$  &Flux$^c$     &L$^{d}$ &$\chi^{2}/$dof    \\
  \noalign{\smallskip}
           &  &           &            &  & C-stat (dof)$^{e}$  \\
 \hline
 \noalign{\smallskip} 
WT (147--1343)  &$2.35_{-0.14}^{+0.15}$  &$0.79_{-0.06}^{+0.06}$  &$1.8$ &1.8 &$681/563$ \\
WT (4625--4697) &$3.18_{-0.65}^{+0.78}$  &$1.13_{-0.27}^{+0.29}$  &$1.3$ &1.4 &$52/40$ \\
PC (6176--6730) &$6.44_{-3.34}^{+4.69}$  &$1.42_{-1.03}^{+1.18}$  &$0.02$ &0.03 &$73.4 (83)$ \\
 \noalign{\smallskip} 
  \hline
  \end{tabular}
  \end{center}
  \begin{list}{}{} 
  \item[$^{\mathrm{a}}$ Spectrum (seconds since trigger). ] 
  \item[$^{\mathrm{b}}$ In units of 10$^{22}$~cm$^{-2}$. ] 
  \item[$^{\mathrm{c}}$ Observed 2--10\,keV fluxes ($10^{-9}$ erg~cm$^{-2}$~s$^{-1}$). ] 
  \item[$^{\mathrm{d}}$ 2--10\,keV luminosities in units of $10^{36}$ erg~s$^{-1}$, at 2.7\,kpc.] 
  \item[$^{\mathrm{e}}$ Cash statistics (C-stat).] 
  \end{list} 
  \end{table} 
%%%%%%%%%%%%%%%%%%%%%%%%%%%%%%%%%%%%%%%%%%%%%%%%%%

%
\begin{table*} %%%%%%%%%%%%%%%%%%%%%%%%%%%%%%%%%%%%%%%%%%%%%%%%%%%%%%%%   TABLE 4
\caption{Best-fit parameters of the phenomenological models  for \src\ and \igr\ broad-band spectra, respectively. In the latter case an absorption edge
was also included. CPL=cut-off powerlaw; HCP=powerlaw multiplied by exponential cut-off with e-folding
factor.}
\label{fits_phenom}
 \begin{tabular}{lrrrr}
\hline
       		          &       \multicolumn{2}{c}{\src}    &       \multicolumn{2}{c}{\igr}   \\
 \hline
Parameter                                   &         CPL 	     &		 HCP	       &	  CPL 	           & HCP \\
\hline

$N_{\rm H} (10^{22}~ \textrm{cm}^{-2})$     & 1.97$^{+0.17}_{-0.16}$ &$1.76^{+0.17}_{-0.15}$   &    0.73$^{+0.06}_{-0.06}$ &$1.01^{+0.06}_{-0.05}$  \\	      
$E_{\rm edge}$ (keV)  		      	    & --		     & --		       & 22.98$^{+  1.15}_{-  1.12}$		   & -- \\
$\tau_{\rm edge}$     		      	    & --		     & --	               & 1.35$^{+0.54}_{-0.43}$    &  --\\
$\Gamma$                              	    &0.25$^{+0.11}_{-0.10}$  &$0.42^{+0.09}_{-0.09}$   & 0.04$^{+0.07}_{-0.07}$    &$0.69^{+0.02}_{-0.04}$  \\
$E_{\rm c}$ (keV)                     	    & --                     &$9.56^{+0.99}_{-0.82}$   &  --                       &$5.66^{+0.54}_{-0.51}$  \\
$E_{\rm f}$ (keV)                     	    &9.19$^{+0.95}_{-0.80}$  &$4.60^{+0.37}_{-0.36}$   & 7.50$^{+0.54}_{-0.49}$    &$14.32^{+0.32}_{-1.61}$  \\
$F^{\rm a}_{\rm 2-10\,keV}$                 & 2.4		     &$2.3$	               & 3.1			   &3.2  \\
$F^{\rm a}_{\rm 1-100\,keV}$   	            & 7.0		     &$7.1$		       & 7.6			   & 9.0 \\
$L^{\rm b}_{\rm 2-10\,keV}$    	            & 2.0		     &$2.0$		       & 4.8			   &5.0  \\
$L^{\rm b}_{\rm 1-100\,keV}$   	            & 6.1		     &$6.1$		       & 11.9			   & 14.1 \\
\chiq/dof                             	    & 364/243  	             &$336/242$		       & 344/296		   &324/297   \\
\hline
   \hline
   \end{tabular}
   \begin{list}{}{}
     \item[$^{\mathrm{a}}$]{Average observed flux in units of $10^{-9}$ erg\,cm$^{-2}$\,s$^{-1}$}. 
    \item[$^{\mathrm{b}}$]{In units of $10^{36}$ erg ~s$^{-1}$}. 
   \end{list}
   \end{table*} %%%%%%%%%%%%%%%%%%%%%%%%%%%%%%%%%%%%%%%%%%%%%%%%%%%%%

%%%%%%%%%%%%%%%%%%%%%%%%%%%%%%%%%%%%%%%%%%%%%%%%%%%%%%%%%
\section{Analysis and Results} \label{results} 
%%%%%%%%%%%%%%%%%%%%%%%%%%%%%%%%%%%%%%%%%%%%%%%%%%%%%%%%% 

Fig.~\ref{fig:xrtlcvs} shows the {\it Swift}/XRT 
0.2--10\,keV light curve of \src\ and \igr\ throughout our 2007--2009 monitoring 
program \citep[][and references therein]{Romano2011:sfxts_paperVI} 
background-subtracted and corrected for pile-up, PSF losses, 
and vignetting. All data in one observation (1--2\,ks) were generally 
grouped as one point, except for outbursts, which show up as a vertical lines on the 
adopted scale. The data already presented elsewhere are in grey. 
The details of the 2011 outbursts of \src\ and \igr, referred to their respective 
BAT triggers, are shown in Fig.~\ref{fig:xrtlcvs2}. 
In both cases the XRT caught the descending portion of a very bright 
flare in great detail. There is a similarity between the two outbursts, 
and also with previous outbursts of these sources 
[see figure 6 of \citet[][]{Romano2011:sfxts_paperVII}, 
for a compendium of the best light curves collected before the currently presented ones
and figure 9b--e of \citet[][]{Sidoli2009:sfxts_paperIV} 
for a comparison of different outbursts of the same source].

The details of the data selection and spectroscopic analysis on both sources 
are reported in the remainder of this Section.  
Here we summarise the models we adopted for the broad-band spectra. 
We first considered the phenomenological models typically used to describe 
the X--ray emission from accreting pulsars in HMXBs, i.e., 
{\it i)} simple absorbed (\wabs\ in \textsc{xspec}) power laws; 
{\it ii)} absorbed power laws with exponential cut-offs (\cpl, hereon CPL); 
{\it iii)} absorbed power laws with high-energy cut-offs (\hcp, hereon HCP). 
The simple advantage of this approach is that these fits yield 
easy-to-compare estimates of fluxes and luminosities. 
The disadvantage, however,  is that little physical insight can be 
obtained from such fits, therefore we also considered the following 
physical models:
{\it iv)} absorbed generic Comptonization models (\comptt) in diffusion approximation 
for a disc geometry without a dynamical bulk component,
{\it v)} a combination of two blackbodies (BB) with different temperatures and radii
(\bb$+$\bb), sometimes used to fit the spectra of magnetars \citep[][]{Israel2008_mn}; 
{\it vi)} the new \compmag\ model, recently developed by F12.
We refer the reader to that paper for a detailed description of the algorithm. 
Here we briefly remind the reader that \compmag\ is based on the solution of the radiative transfer equation
for the case of cylindrical accretion onto the polar cap of a magnetized NS.
The velocity field of the accreting matter can be increasing towards the NS surface, or it 
may be described by an approximate decelerating profile.
In the former case, the free parameters are the terminal velocity at the NS surface, $\b0$, and
the index of the law $\beta(z) \propto z^{-\eta}$, while in the second case the law
is given by $\beta(\tau) \propto -\tau$.
The other free parameters of the model are the temperature of the BB seed photons, $\ktbb$,
the electron temperature and vertical optical depth of the Comptonization plasma $\kte$ and $\tau$, 
respectively, and the radius of the accretion column $r_{\rm 0}$, in
units of the NS \scw\ radius.
The different combinations of these parameters determine the steepness of the spectrum
at high energies and the rollover energy position.

However, given that the Comptonization spectra can be determined by only three quantities
(slope, cut-off position and normalisation), for practical purposes it is generally
not possible to keep free all the above mentioned parameters, a procedure which
otherwise lead them to be completely unconstrained during the fitting procedure. 
In the following Sections we will report in more detail the procedure that
we adopted in the spectral fitting.

       %%%%%%%%%%%%%%%%%%%%%%%%%%%%%%%%%%%%%%%%%%%%%%%%%%%%%%%%%
       \subsection{XTE~J1739$-$302} \label{spec_out17391}
       %%%%%%%%%%%%%%%%%%%%%%%%%%%%%%%%%%%%%%%%%%%%%%%%%%%%%%%%%
The XRT light curve (Fig.~\ref{fig:xrtlcvs2}a) shows the descending part of a flare, 
that started off at a peak exceeding 30\,counts s$^{-1}$ then decreases to about 
2.5\,counts s$^{-1}$. A second flare is  observed in the second orbit, with a 
count rate in the range 10--17\,counts s$^{-1}$. 
In the third orbit the count rate was $\sim 0.2$\,counts s$^{-1}$. 
In Fig.~\ref{fig:17391lcv_allbands} are reported the first orbit data of \src\ in several energy bands. 
The event-by-event mask-weighted light curves only show the initial flare, 
which started at $\la T-70$\,s with a slow rise, well defined in both soft bands. 
The source was still bright when the BAT event data end, at $T+900$\,s. 
The BAT survey data cover the same time-span. 

We extracted an XRT spectrum for each orbit of data 
as detailed in Table~\ref{tab:xrtband_17391} 
and we fit them with an absorbed power-law model.  
The results are summarized in Table~\ref{tab:xrtband_17391}.
By linking the $N_{\rm H}$ across the spectra, we can detect 
$N_{\rm H}$ variations, while the photon index is consistent with being
constant, as previously reported in 
\citet{Sidoli2009:sfxts_paperIII,Sidoli2009:sfxts_paperIV}.

The BAT average spectrum ($T-119$ to T$+903$\,s) can be fit by a simple power-law model
with a photon index $\Gamma=2.67_{-0.15}^{+0.16}$ ($F_{\rm 15-50\,keV}=2.8\times10^{-9}$ 
erg cm$^{-2}$ s$^{-1}$); the fit, however, is not good 
(\chiq/dof=51/37) and the residuals indicate that a curvature is present in the
spectrum, which is best assessed in a broad-band analysis.

\begin{figure}
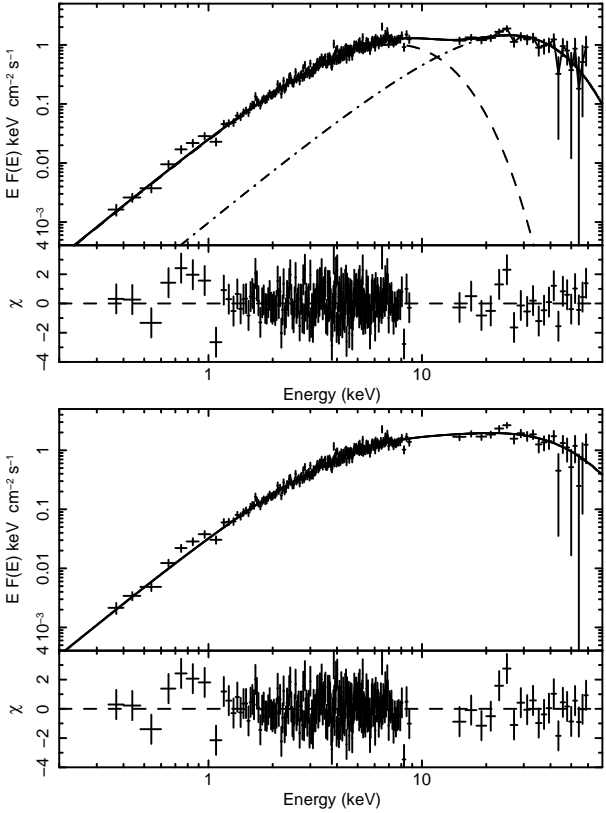
 %%%%%%%%%%%%%%%%%%%%%%%%%%%%%%%%%%%%%%%%%%%%%%%%%%%%%   FIGURE 4
\begin{center}
\vspace{-0.5truecm}
\hspace{+1.0truecm}
\centerline{\includegraphics[angle=270,width=8.0cm]{figure4a.ps}}
\centerline{\includegraphics[angle=270,width=8.0cm]{figure4b.ps}}
\end{center}
\caption{Absorption-corrected unfolded EF(E) models and spectra, and residuals between the data and
the model in units of $\sigma$ for \src. {\it Upper panel}: \bb$+$\bb. {\it Lower panel}: \compmag.}
\label{eeuf_1739}
\end{figure}%%%%%%%%%%%%%%%%%%%%%%%%%%%%%%%%%%%%%%%%%%%%%%%%%%%%%

\begin{table} %%%%%%%%%%%%%%%%%%%%%%%%%%%%%%%%%%%%%%%%%%%%%%%%%%%%%%%%   TABLE 5
\caption{Best-fit parameters of the model \wabs $\times$(\bb+\bb)  for \src\ and \igr\ broad-band spectra, respectively.}
\label{fits_2bb}
 \begin{tabular}{lll}
 \hline
 Parameter                            & \src                    & \igr\  \\
\hline
$N_{\rm H}~ (10^{22}~ \textrm{cm}^{-2})$  & 1.07$^{+0.09}_{-0.09}$  & 0.58$^{+0.07}_{-0.06}$ \\
$\ktbb^1$ (keV)                       & 1.88$^{+0.07}_{-0.06}$  & 0.93$^{+0.07}_{-0.07}$ \\
$R_{\rm{bb},1}$ (km)                  & 1.15$^{+0.04}_{-0.05}$                     & 3.43$^{+0.41}_{-0.34}$\\
$\ktbb^2$   (keV)                    & 6.62$^{+0.49}_{-0.15}$  & 3.45$^{+0.08}_{-0.07}$ \\
$R_{\rm{bb},2}$ (km)                  & 0.11$^{+0.01}_{-0.01}$                     & 0.77$^{+0.04}_{-0.04}$ \\
$F^{\rm a}_{\rm 2-10\,keV}$     & 2.2                     & 3.1 \\
$F^{\rm a}_{\rm 1-100\,keV}$    & 5.9                     & 7.7\\
$L^{\rm b}_{\rm 2-10\,keV}$    & 1.9                     & 4.8\\
$L^{\rm b}_{\rm 1-100\,keV}$   & 5.2                     & 12.1 \\
\chiq/dof                             & 285/241             &  335/297 \\
\hline
\end{tabular}
   \begin{list}{}{} 
    \item[$^{\mathrm{a}}$]{In units of $10^{-9}$ erg\,cm$^{-2}$\,s$^{-1}$}. 
    \item[$^{\mathrm{b}}$]{In units of $10^{36}$ erg~ s$^{-1}$}. 
   \end{list}
\end{table}%%%%%%%%%%%%%%%%%%%%%%%%%%%%%%%%%%%%%%%%%%%%%%%%%%%%%
\begin{table} %%%%%%%%%%%%%%%%%%%%%%%%%%%%%%%%%%%%%%%%%%%%%%%%%%%%%%%%   TABLE 6
\caption{Best-fit parameters of the model \wabs $\times$\comptt\ for \src\ and \igr\ broad-band spectra, respectively. As for the case
of Table \ref{fits_phenom}, an absorption edge was included in the model for \igr.} %\ref{fits_cpl}
\label{fits_comptt}
 \begin{tabular}{lll}
 \hline
 Parameter                            & \src                      & \igr\  \\
\hline
$N_{\rm H} ~(10^{22}~ \textrm{cm}^{-2})$                           & 0.81$^{+0.11}_{-0.10}$ & 0.58$^{+0.07}_{-0.06}$ \\
$E_{\rm edge}$ (keV)                  & --                        & 20.97$^{+1.82}_{-1.23}$ \\
$\tau_{\rm edge}$                     & --                        & 0.62$^{+0.35}_{-0.31}$ \\
$\ktw$ (keV)                          & 1.34$^{+0.07}_{-0.06}$    & 0.63$^{+0.07}_{-0.08}$ \\
$\kte$ (keV)                               & 8.83$^{+1.88}_{-1.11}$    & 3.92$^{+0.18}_{-0.16}$ \\
$\tau$                                & 3.61$^{+0.51}_{-0.79}$    & 10.67$^{+0.71}_{-0.74}$  \\
$F^{\rm a}_{\rm 2-10\,keV}$    & 2.2                       &  3.0\\
$F^{\rm a}_{\rm 1-100\,keV}$   & 6.8                       & 8.0\\
$L^{\rm b}_{\rm 2-10\,keV}$   & 1.9                       & 4.7\\
$L^{\rm b}_{\rm 1-100\,keV}$  & 5.9                       &  12.5\\
\chiq/dof                             & 289/242               &  323/295               \\
\hline
\end{tabular}
   \begin{list}{}{} 
    \item[$^{\mathrm{a}}$]{In units of $10^{-9}$ erg\,cm$^{-2}$\,s$^{-1}$}. 
    \item[$^{\mathrm{b}}$]{In units of $10^{36}$ erg~s$^{-1}$}. 
   \end{list}
\end{table}%%%%%%%%%%%%%%%%%%%%%%%%%%%%%%%%%%%%%%%%%%%%%%%%%%%%%
\begin{table} %%%%%%%%%%%%%%%%%%%%%%%%%%%%%%%%%%%%%%%%%%%%%%%%%%%%%%%%   TABLE 7
\caption{Best-fit parameters of the model \wabs $\times$\compmag\  for \src\ and \igr\ broad-band spectra, respectively.
In both cases the fixed parameters are $\eta=0.5$, $\b0=0.05$, $r_{\rm 0}=0.25$, and $A=1$. Also in this case,
as in Table \ref{fits_phenom}, an absorption edge was included in the model for \igr. } %\ref{fits_cpl} 
\label{fits_compmag}
 \begin{tabular}{lll}
 \hline
 Parameter            & \src                 & \igr\  \\
\hline
$N_{\rm H}~(10^{22}~ \textrm{cm}^{-2}) $                           & 1.18$^{+0.12}_{-0.09}$    & 0.68$^{+0.10}_{-0.08}$ \\
$E_{\rm edge}$ (keV)                  & --                        &  21.71$^{+  1.84}_{-  0.95}$\\
$\tau_{\rm edge}$                     & --                        & 0.81$^{+0.52}_{-0.23}$ \\
$\ktbb$ (keV)                         & 1.63$^{+0.10}_{-0.17}$    & 0.73$^{+0.11}_{-0.10}$ \\
$\kte$  (keV)                           & 9.62$^{+4.05}_{-2.57}$    & 3.47$^{+0.34}_{-0.02}$ \\
$\tau$                                & 0.33$^{+0.30}_{-0.03}$    & 1.19$^{+0.09}_{-0.10}$ \\
$F^{\rm a}_{\rm 2-10\,keV}$    & 2.3                       & 3.0\\
$F{\rm a}_{\rm 1-100\,keV}$   & 7.4                       & 8.1\\
$L^{\rm b}_{\rm 2-10\,keV}$    & 1.9                       & 4.7\\
$L^{\rm b}_{\rm 1-100\,keV}$   & 6.4                       &  12.7\\
\chiq/dof                             & 292/241          &  324/295 \\
\hline
\end{tabular}
   \begin{list}{}{} 
    \item[$^{\mathrm{a}}$]{In units of $10^{-9}$ erg\,cm$^{-2}$\,s$^{-1}$}. 
   \item[$^{\mathrm{b}}$]{In units of $10^{36}$ erg~s$^{-1}$}. 
   \end{list}
\end{table}%%%%%%%%%%%%%%%%%%%%%%%%%%%%%%%%%%%%%%%%%%%%%%%%%%%%%

In order to perform broad-band spectroscopy of the outburst, we extracted strictly 
simultaneous spectra from the XRT and BAT event lists, i.e., in the 
time interval 147--903\,s since the BAT trigger (see Fig.~\ref{fig:17391lcv_allbands}).
We fit them in the 0.5--10\,keV and 15--60\,keV
energy bands for XRT and BAT, respectively. 
Factors were included in the fitting to allow for normalisation 
uncertainties between the two instruments, constrained within their 
usual ranges (0.9--1.1). 
A simple absorbed power-law model clearly yields an inadequate fit  
of the broad-band spectrum with a \chiq/dof=926/244. 
A significant improvement is obtained when considering  
the other phenomenological curved models described above (see  
Table~\ref{fits_phenom}).

A more satisfactory fit is instead obtained with the model  \wabs$\times$(\bb+\bb),
whose results are reported  in Table \ref{fits_2bb},
while in Fig. \ref{eeuf_1739} we show the deconvolved best-fit model and data.

We also tested the model \wabs$\times$\comptt, which yields 
statistically similar results, and we report
the main values of the model in Table \ref{fits_comptt}.

Finally, we tested the \compmag\ model assuming a free-fall like velocity profile ($\eta=0.5$), 
two terminal velocities at the NS surface, $\b0=0.05$ and $\b0=0.2$, respectively, and
 NS albedo $A=1$.
The free parameters of the model during the fitting procedure were the seed BB temperature $\ktbb$,
the electron temperature $\kte$, and the vertical optical depth of the accretion
column $\tau$. 
We note that the latter quantity in this case is a factor about 1/1000 lower than classical
 optical depth because of the inclusion of an energy-independent correction
term which reduces the Thomson cross-section $\sigma_{\rm T}$ for photons 
propagating in the direction of 
the magnetic field \citet[][hereafter BW07]{BeckerWolff2007}. 

We initially set the radius of the accretion column $r_{\rm 0}=0.25$, which corresponds to 
$\sim 1$ km for a NS star with mass $M=1.4 M_{\odot}$; the results for the case
$\b0=0.05$ are reported in Table \ref{fits_compmag}, while in Fig. \ref{eeuf_17544}
we show the deconvolved best-fit model and data.

No significant variation is observed in the \chiq-value when increasing  $r_{\rm 0}$  from 0.25 to 1.
A similar statistical result is also obtained assuming 
$\b0=0.2$ (\chiq/dof=294/241). In this case, a slight decrease is observed both
in the electron temperature and optical depth, although their values can be considered
unchanged within errors ($\kte=6.5^{+ 8.5}_{-0.8}$ keV and $\tau=0.29^{+0.11}_{-0.01}$,
respectively).
The electron temperature drop is not surprising, as in a thermal plus bulk Comptonization scenario the spectral slope and
cut-off energy are dictated not only by the plasma temperature and optical depth, but also by the shape of the velocity
field and maximum velocity of the accreting matter (see fig.~4 in F12).
Thus, for given (observed) spectral slope and roll-over energy, the higher the values of $\b0$, the lower the values of
$\kte$, as the bulk Comptonization progressively increases its importance at expenses of thermal Comptonization.
Even in this case, with $\b0=0.2$, the \chiq\ remains unchanged for varying $r_{\rm 0}$ from 0.25 up to 1.

An important issue we want to emphasize is that for \src\ the value of $\tau$ obtained with \compmag\ 
is less than 1 (see Table \ref{fits_compmag}), and actually the higher the assumed value of $\b0$ the lower the value of $\tau$.

As the \compmag\ model is based on the solution of the Fokker-Planck approximation of the radiative 
transfer equation, it generally holds for an optical depth $\tau \ga 1$. 
Our fits on \src\ data yield values of $\tau$ which are formally smaller than unity. 
In general terms, this would indicate that 
the diffusion along the direction of the column axis is not efficient, hence 
either the spatial diffusion of the photons is only efficient for 
photons propagating perpendicularly with respect to the column axis, 
or the system is significantly anisotropic due to the presence of the magnetic field. 
In either case, the diffusion approximation would not be recommended, as inadequately 
describing the system (further details can be found in the Appendix). 

We note however that, in our case, an acceptable fit can be achieved with 
a value of the optical depth within a factor of 3 of unity, so that the obtained
physical parameters can be discussed with reasonable accuracy. In particular, the
electron temperature $\kte$ results consistent within errors with that obtained
with \comptt\ and with the hotter BB temperature in the two-BB model (see Tables \ref{fits_2bb}
and \ref{fits_comptt}).

       %%%%%%%%%%%%%%%%%%%%%%%%%%%%%%%%%%%%%%%%%%%%%%%%%%%%%%%%%
       \subsection{IGR~J17544$-$2619} \label{spec_out17544}
       %%%%%%%%%%%%%%%%%%%%%%%%%%%%%%%%%%%%%%%%%%%%%%%%%%%%%%%%%

%%%%%%%%%%%%%%%%%%%%%%%%%%%%%%%%%%%%%%%%%%%%%%%%%%   TABLE 4
 \begin{table}
\tabcolsep 3pt   
 \begin{center}
 \caption{Absorbed power-law spectral fits of XRT data of \igr.}
 \label{tab:xrtband_17544}
 \begin{tabular}{lllllll}
% \hline
 \hline
 \noalign{\smallskip}
 Spectrum$^a$      &$N_{\rm H}^b$  &$\Gamma$  &Flux$^c$     &L$^{d}$ &$\chi^{2}/$dof    \\
  \noalign{\smallskip}
           &  &           &            &  & C-stat (dof)$^{e}$  \\
 \hline
 \noalign{\smallskip} 
WT (133--783)  &$1.00_{-0.06}^{+0.06}$ &$0.69_{-0.04}^{+0.04}$   &$3.0$ &4.89 & $545/570$ \\
PC (784--1659) &$1.83_{-0.49}^{+0.55}$ &$1.05_{-0.27}^{+0.28}$   &$0.27$ &0.46 &$30/33$ \\
PC (4881--7418) &$2.33_{-0.41}^{+0.47}$ &$1.39_{-0.23}^{+0.25}$  &$0.06$ &0.11 &$46/42$ \\
  \noalign{\smallskip} 
  \hline
  \end{tabular}
  \end{center}
  \begin{list}{}{} 
  \item[$^{\mathrm{a}}$ Spectrum (seconds since trigger). ] 
  \item[$^{\mathrm{b}}$ In units of 10$^{22}$~cm$^{-2}$. ] 
  \item[$^{\mathrm{c}}$ Observed 2--10\,keV fluxes ($10^{-9}$ erg~cm$^{-2}$~s$^{-1}$). ] 
  \item[$^{\mathrm{d}}$ 2--10\,keV luminosities in units of $10^{36}$ erg~s$^{-1}$, at 3.6\,kpc.] 
  \item[$^{\mathrm{e}}$ Cash statistics (C-stat).] 
  \end{list} 
  \end{table} 
%%%%%%%%%%%%%%%%%%%%%%%%%%%%%%%%%%%%%%%%%%%%%%%%%%

%%%%%%%%%%%%%%%%%%%%%%%%%%%%%%%%%%%%%%%%%%%%%%%%%%   FIGURE 5
\begin{figure}
\begin{center}
\vspace{-0.5truecm}
\hspace{+1.0truecm}
\centerline{\includegraphics*[angle=270,width=9.0cm]{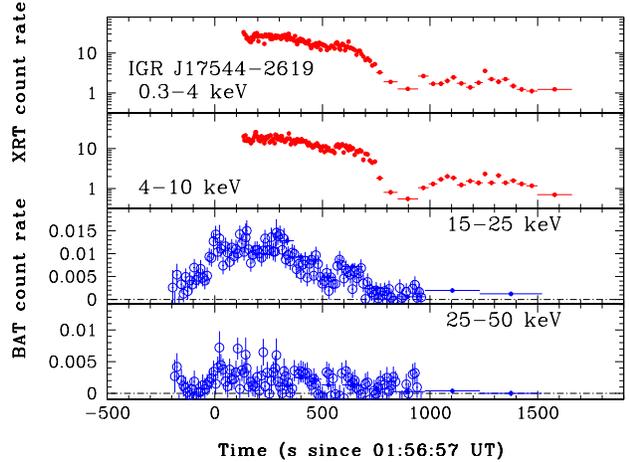}}
\end{center}
\caption{Same as Fig.~\ref{fig:17391lcv_allbands} for 
the 2011 March 24 outburst of \igr. The empty circles correspond to BAT in event mode,  
filled circles to BAT survey mode data. 
}
\label{fig:17544lcv_allbands}
\end{figure}
%%%%%%%%%%%%%%%%%%%%%%%%%%%%%%%%%%%%%%%%%%%%%%%%%%

The XRT light curve of \igr\ is reported in Fig.~\ref{fig:xrtlcvs2}b, 
and shows the two bright flares observed in the first orbit; 
the first was caught (WT mode) in the descending part and reached a peak exceeding 
50\,counts s$^{-1}$; the second (PC mode) started off at about $T+750$\,s and reached 
$\sim 5$ counts s$^{-1}$. At about $T+5000$\,s (second orbit) the source is 
still at $\sim 1$\,counts s$^{-1}$. 
Fig.~\ref{fig:17544lcv_allbands} shows the initial data of \igr\ in 
several energy bands. 
The BAT event-by-event mask-weighted light curves show a slow rise from $T-200$\,s 
up to the first peak at $T-0$\,s, followed by a few more flares, then 
follows the decaying shape of the XRT light curve. 
The BAT survey data cover a longer time-span, and show that the 
source was still detected in the soft band out to $T+1500$\,s.

We extracted three XRT spectra at different times since the trigger and
we fit them with an absorbed power-law model, as detailed in
 Table~\ref{tab:xrtband_17544}.  
The results are summarized in Table~\ref{tab:xrtband_17544}.
By linking the $N_{\rm H}$ across the spectra, we can detect 
a harder-when-brighter trend and $N_{\rm H}$ variations, 
as previously observed by \citet{Rampy2009:suzaku17544}.
We note that this is the brightest outburst of this source recorded by \sw.

The BAT average spectrum ($T-239$ to T$+963$\,s) was fit by a simple power-law model
with a photon index $\Gamma=4.19_{-0.32}^{+0.33}$ ($F_{\rm 15-50\,keV}=2.4\times10^{-9}$ 
erg cm$^{-2}$ s$^{-1}$, \chiq/dof=22/20). 

We extracted strictly simultaneous spectra from the XRT and BAT event lists
(133--783\,s since the BAT trigger; see Fig.~\ref{fig:17544lcv_allbands}) and 
we fit them in the 0.5--10\,keV and 15--60\,keV energy bands for XRT and BAT, respectively,
by adopting the same models as those used for \src. 
Factors were included in the fitting to allow for normalisation 
uncertainties between the two instruments, constrained within their 
usual ranges (0.9--1.1). 

As is the case of \src, for \igr\ a simple absorbed power-law model is inadequate to 
fit the broad-band spectrum (\chiq/dof=2071/299). 
A significant improvement is obtained when considering 
the CPL and HCP models (Table~\ref{fits_phenom}).

\begin{figure}
\begin{center}
\centerline{\includegraphics[angle=-90,width=10cm]{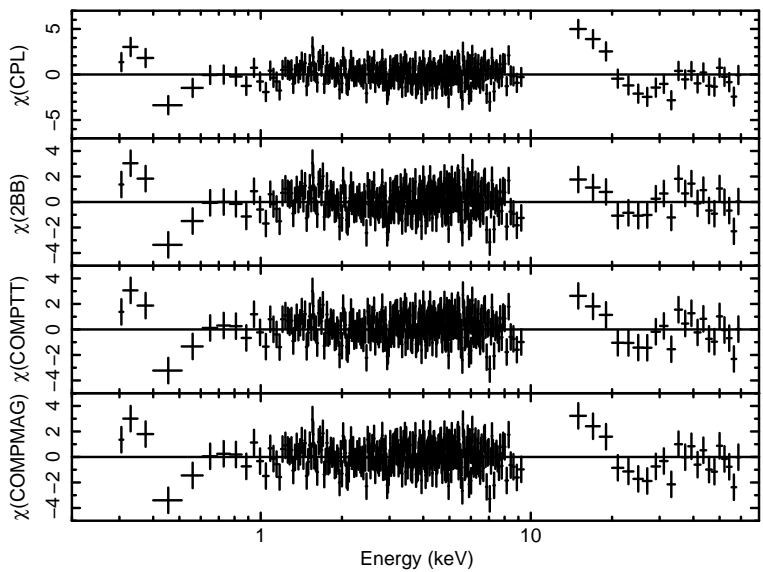}}
\end{center}
\caption{Residuals to data in units of $\sigma$ of the four  models adopted to fit the continuum of \igr\ (see Sect.~\ref{spec_out17544} for details).}
\label{residuals_17544}
\end{figure}%%%%%%%%%%%%%%%%%%%%%%%%%%%%%%%%%%%%%%%%%%%%%%%%%%%%% 

The main difference in the \chiq-values between the CPL and HCP models is
due to the residuals of the BAT spectrum above 30 keV, which show
in the former case  a  sinusoidal-like feature in the 15--30 keV region
(see Fig. \ref{residuals_17544}).
The inclusion of an absorption edge in the CPL model improves the statistical result from \chiq/dof=383/298 
to \chiq/dof=344/296, with best-fit parameters reported in Table \ref{fits_phenom}.
The F-test for discriminating among two different
models (namely without and with the absorption edge) provides however a
probability of chance improvement (PCI) of about only 20\,\%.
The fact that these residuals are not observed in the HCP model besides on fact
that in this case the spectrum is less smooth due to the presence of the
$E_{\rm c}$ term. The well constrained determination of the latter parameter
is actually indicative for the presence of a change of curvature in
the \mbox{X-ray} spectrum around 15--20 keV.

The models \wabs $\times$ \comptt\ yields \chiq/dof=339/297, and the inclusion of
the absorption edge improves the fit to  \chiq/dof=323/295, with however an even
lower PCI ($\sim$ 35\%) with respect to the case of a CPL, and best-fit parameters
reported in Table \ref{fits_comptt}.

Similarly to \src, a satisfactory fit can be obtained by the sum of two blackbody spectra (see Table \ref{fits_2bb}), 
and in this case the inclusion of the absorption edge {\bf only marginally} improves the fit  by $\Delta$\chiq $\sim$ 7.

Performing the spectral analysis with the \compmag\ model, we followed the
same procedure of \src.
First, we considered the case of free-fall velocity profile fixing $\b0=0.05$, accretion column
radius $r_{\rm 0}=0.25$ and NS albedo $A=1$.
In this case we obtain \chiq/dof=343/297, with an improvement
$\Delta$\chiq$\sim$ 20 when including the absorption edge at 20 keV.
Also for \igr,  the fit results to be insensitive to $r_{\rm 0}$, with the \chiq-value substantially
unchanged for $r_{\rm 0}$ increasing from 0.25 to 1.

In the second case, namely $\b0=0.2$, the inclusion of the absorption edge reduces similarly the fit 
by $\Delta$\chiq$\sim$ 20, but we observed a decreases from \chiq/dof=363/295 to   \chiq/dof=332/295 by setting
$r_{\rm 0}=0.25$ and $r_{\rm 0}=1$, respectively.
For higher values of $r_{\rm 0}$, no further improvement of the \chiq\ was observed, so that we consider it as our
actual best-fit parameter.

Unlike the case of \src, the increased value of  $\b0$ leads to 
a more significant decrease of the electron
temperature, with  $\kte= 1.2^{+0.1}_{-0.1}$ keV, while 
the optical depth to first approximation remains constant within
errors, with $\tau=1.44^{+0.11}_{-0.09}$.

The reason for this is that a higher value of $\b0$ is compensated
by the simultaneous increase of the accretion column radius $r_{\rm 0}$ from 0.25 to 1.

\begin{figure}%%%%%%%%%%%%%%%%%%%%%%%%%%%%%%%%%%%%%%%%%%%%%%%%%%%%%   FIGURE 7 
\begin{center}
\vspace{-0.5truecm}
\hspace{+1.0truecm}
\centerline{\includegraphics[angle=270,width=8.0cm]{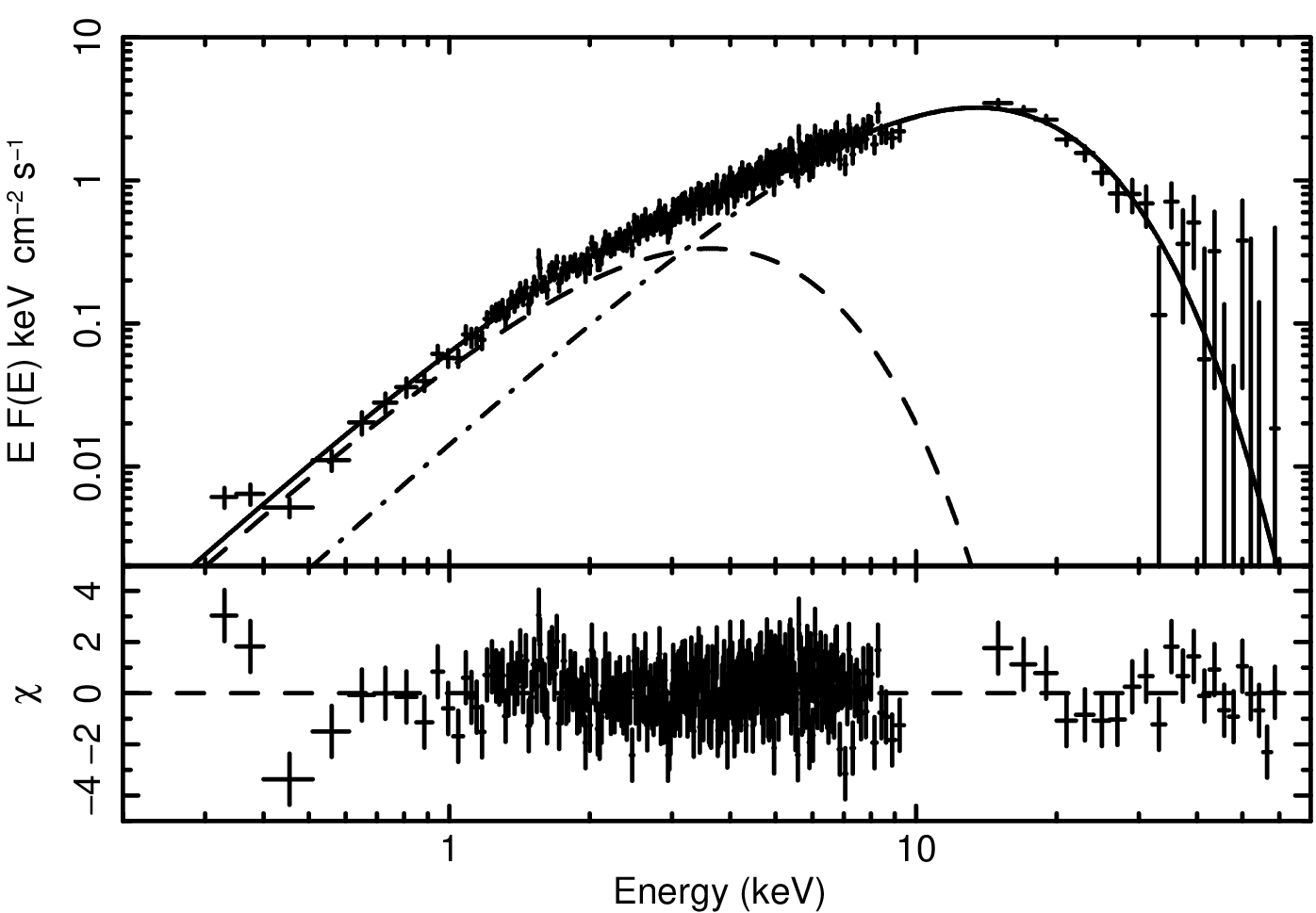}}
\centerline{\includegraphics[angle=270,width=8.0cm]{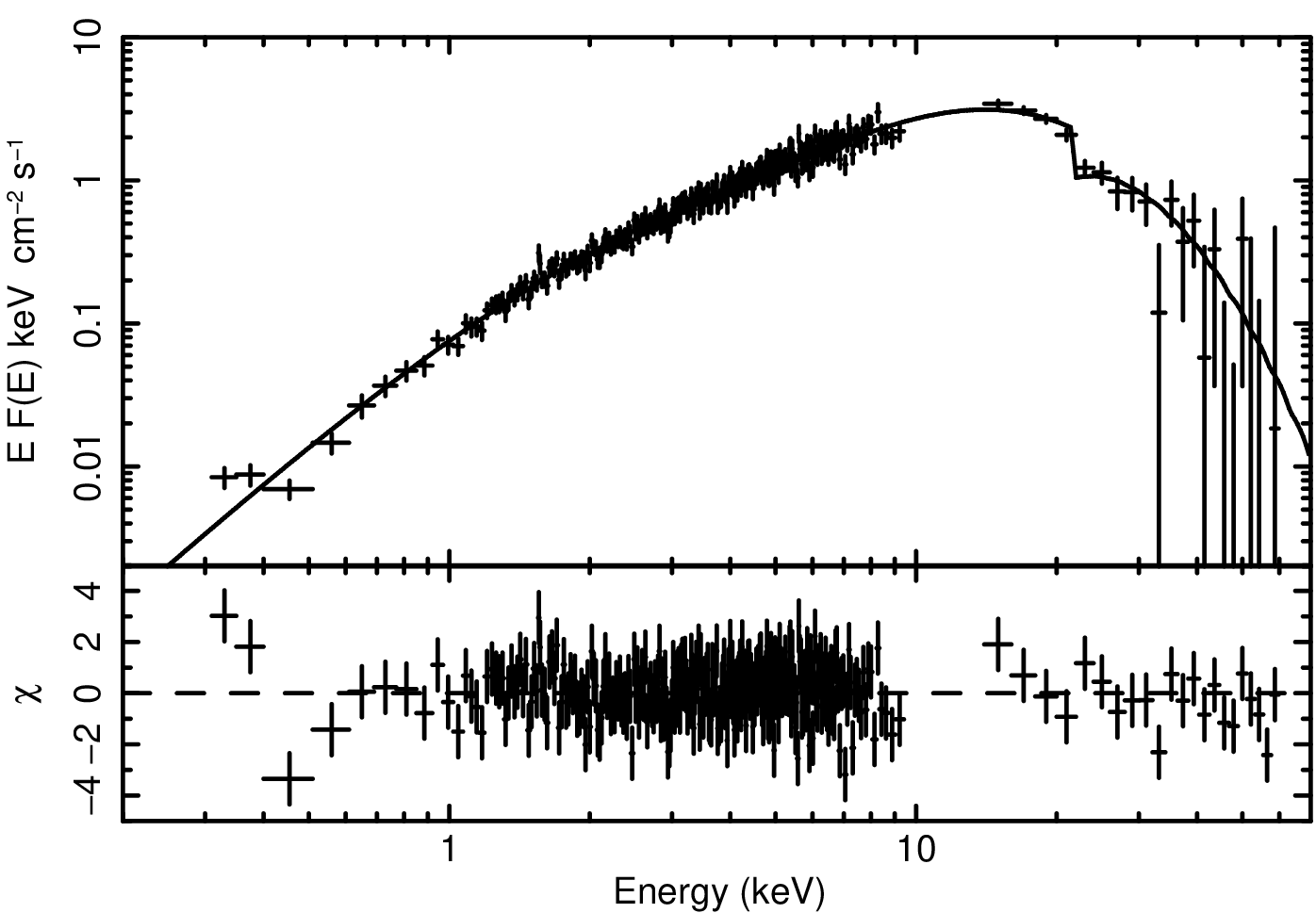}}
\end{center}
\caption{Asborption-corrected EF(E) spectra, best-fit models and residuals between the data and
the model in units of $\sigma$ for \igr. 
{\it Upper panel}: \bb$+$\bb. 
{\it Lower panel}: \edge $\times$\compmag.}
\label{eeuf_17544}
\end{figure}

In Fig.~\ref{residuals_17544} we report as a summary,  the residuals to the data in units of $\sigma$ 
of the four different adopted models above described, while Fig.~\ref{eeuf_17544} shows the
 absorption-corrected best-fit models and residuals between the data and
the model in units of $\sigma$ for \bb$+$\bb\ and \edge $\times$\compmag.

%%%%%%%%%%%%%%%%%%%%%%%%%%%%%%%%%%%%%%%%%%%%%%%%%%%%% 

%%%%%%%%%%%%%%%%%%%%%%%%%%%%%%%%%%%%%%%%%%%%%%%%%%%%%%%%%
\section{Discussion\label{discussion}}
%%%%%%%%%%%%%%%%%%%%%%%%%%%%%%%%%%%%%%%%%%%%%%%%%%%%%%%%% 

\src\ and \igr\ are considered the prototypes of the SFXT class and, as such,  
were systematically observed with \sw, which caught several outbursts from both sources. 
The simplest way to parametrise the shape of the broad-band spectra is to 
fit them with phenomenological models typically used to 
describe the X--ray emission from accreting pulsars in HMXBs, i.e., 
absorbed power laws,  
absorbed power-law models with a high energy cut-offs and 
absorbed power-law models with exponential cut-offs.  

The advantage of this approach is that the fits yield easy-to-compare 
estimates of fluxes and luminosities for each outburst. 
Indeed, the data presented in this paper yield values of both high energy 
cut-off $E_{\rm c}$ and e-folding energy $E_{\rm f}$ 
(see Sect.~\ref{spec_out17391} and \ref{spec_out17544})
consistent with the ones obtained previously for both sources 
\citep{Sidoli2009:sfxts_paperIII,Sidoli2009:sfxts_paperIV,Romano2011:sfxts_paperVI,Romano2011:sfxts_paperVII,Romano2010:TEXAS10}. 
The disadvantage is, clearly, that little physical insight can be obtained from such fits.
In this paper we also apply the following more physically-motivated models:  
absorbed Comptonization models (\comptt); 
a combination of two blackbodies with different temperatures and radii; 
and for the first time to SFXTs, the new \compmag\ model, recently developed by F12.

There is an interesting aspect to discuss, which has not been yet fully 
faced in the accretion physics of SFXTs.
If we focus on the results using the \comptt\ and \compmag\ models,
it can be shown that the electron density where Comptonization
takes place is of the order of $10^{19}$ cm$^{-3}$. Indeed, the Thomson optical depth is given by
\begin{equation}
\tau \approx 7~n_{\rm 19}~r_{\rm 6},
\end{equation}

\noindent 
where $n_{\rm 19} \equiv n_{\rm e}/10^{19}$  and $r_{\rm 6} \equiv R/10^6$ are
dimensionless electron  density and system length scale, respectively.
If spectral formation occurs close to the NS ($r_{\rm 6} \sim$ 1--2), from
the \comptt\ best-fit values of $\tau$ reported in Table~\ref{fits_comptt} 
we obtain $n_{\rm 19} \sim 1$.

The possible presence of a strong magnetic field ($B \ga 10^{12}$ G), with associated reduction
of the Thomson cross-section $\sigma_{\rm T}$ would actually require an even higher electron
density, which somewhat compensates the lower cross-section value in order to
maintain the same Comptonization parameter $Y$, and in turn the observed spectral index.

This qualitative effect can be tested from the results of the \compmag\ model (Table \ref{fits_compmag}). 
The electron density for column accretion, which is assumed in \compmag,
is given by
\begin{equation}
 n_{\rm e} \approx 10^{19} \frac{\dot{m}}{m r_{\rm 0}^2 \beta_{\rm z}}~ \textrm{cm$^{-3}$},
\label{ne_column}
\end{equation}

\noindent 
where $\dot{m} \equiv \dot{M}/\dot{M}_{\rm Edd}$ is the accretion rate in Eddington units, $m=M_{\rm NS}/M_{\odot}$ is
the NS mass in units of solar masses, $r_{\rm 0}=R_{\rm 0}/(R^{\rm scw}_{\odot} m)$ is the accretion column radius
in units of the NS \scw\ radius, and $\beta_{\rm z}=V_{\rm z}/c$ is the accretion column velocity.
We derived $\dot{m}$ from the best-fit value of $\tau$ reported in Table \ref{fits_compmag} for \igr\ and \src\ 
(with $r_{\rm 0}=0.25$ and $\b0=0.05$) and using equation (44) in F12 for both sources.
Then, substituting $\beta_{\rm z}$ in equation (\ref{ne_column})  with a value averaged 
over the vertical $z$-coordinate $<\beta_{\rm z}>$,
we obtain  $n_{\rm e} \sim 10^{21}$ cm$^{-3}$ in both sources, which is actually a factor about one hundred higher
than that inferred from \comptt.

Then, we consider the mass flow rate across the magnetosphere

\begin{equation} \label{eq. mass flow rate}
\dot{M} = 4 \pi R_{\rm m}^2 \rho v_{\rm in},
\label{mdot_mag}
\end{equation}
where $R_{\rm m}$ is the magnetospheric radius, and $v_{\rm in}$ is the 
infall velocity at $R_{\rm m}$, that is expected to be less than the free-fall velocity 

\begin{equation}
v_{\rm ff} = \left ( \frac{2GM_{\rm NS}}{R} \right )^{1/2}.
\label{v_ff}
\end{equation}
The accretion luminosity at the NS surface on the other hand is

\begin{equation} \label{accretion luminosity}
 L_{\rm acc} \la \frac{GM_{\rm NS}\dot{M}}{R_{\rm NS}}.
 \label{l_acc}
 \end{equation}
Assuming a NS with mass $M=1.4 M_{\odot}$ and radius $R_{\rm NS}=10$ km, and combining  equations (\ref{mdot_mag}) and (\ref{l_acc}) we obtain

\begin{equation}
 \ne \ga 1.3 \times 10^{14} \frac{L_{\rm 37}}{r^2_8 v_9}~ \textrm{cm$^{-3}$},
\label{ne_mag}
\end{equation}
where $L_{\rm 37} \equiv L_{\rm x}/10^{37}$, $r_8 \equiv r_{\rm m}/10^8$, and $v_9 \equiv v_{\rm in}/10^9$,
respectively.

We consider a magnetospheric radius  $r_{\rm m} \approx 10^8$\,cm, calculated assuming the typical values of the magnetic field
of an accreting pulsar in HMXBs and the typical stellar wind parameters of OB supergiants \citep{do73}.
Additionally, for the infall velocity we assume for sake of simplicity $v_{\rm in} \approx v_{\rm ff}$.

With these prescriptions in mind and noting that  both sources during the joint XRT/BAT observation have shown $L_{\rm x} \la 10^{37}$ erg~s$^{-1}$,
from equation (\ref{ne_mag}) we find that at the magnetospheric radius $\ne \ga 10^{14}$ cm$^{-3}$.

This value of the electron density at the magnetospheric radius is about seven orders of magnitude lower
than that inferred from the best-fit parameters of \compmag\ as shown above, and 
thus cannot be simply a result of a simplified treatment of the problems such as pure spherical accretion
assumed in  equation (\ref{ne_mag}).

However, there are at least to issues which are worth mentioning.
First, the electron density estimated from the best-fit parameters of \compmag\ is related to
the region where the \mbox{X-ray} spectral formation is assumed to take place, which, in the case
of \compmag, correspond to cylindrical accretion column close to the NS surface with
characteristic height-scale $H \approx$ 1-2 $R_{\rm NS}$.

This scale-length is about 1/100 of the assumed magnetospheric radius ($\approx 100 R_{\rm ns}$),
and for a density scaling as $R^{-2}$ it would imply to first approximation 
a density increase from $R_{\rm m}$ $R_{\rm NS}$ about $\ne^{\rm NS}/\ne^{\rm m} \sim 10^4$.
Of course, matter channeling towards the magnetic poles may also  play an important role
in increasing the matter density in the region close to the NS surface.

The second point to be considered is that if the NS orbital motion is supersonic,
 strong shock waves are expected to form approximatively at the NS magnetosphere
(bow shock region). The net result would be thus an increase of the electron density
with respect to the value reported in equation (\ref{ne_mag}).

The Mach number of the NS is given by
\begin{equation}
 \mathcal{M}  \approx 10^2 V^{\rm ns}_8 (\gamma Z T_{\rm e}/\mu)^{-1/2},
\label{mach}
\end{equation}

\noindent 
where $V^{\rm ns}_8$ is the NS orbital velocity in units of $10^{8}$ cm~s$^{-1}$, $\gamma$ is the adiabatic index
of the wind, $\mu=m_{\rm i}/m_{\rm p}$ its molecular weight, Z its the charge state, and $T_{\rm e}$ its  temperature in eV.
For wind temperatures in the range $10^5-10^7$ K \citep{Ducci2009} , and  velocities of the NS derived
from the orbital parameters for both for \src\ and \igr, $\mathcal{M}$ varies from a few to about 30, 
 and if the shock is isothermal, $\rho_1/\rho_2 \approx \mathcal{M}^2$. Actually, a density increase
of a factor $\sim 10^2$ at the magnetospheric shock front, with respect to the value computed 
in equation (\ref{ne_mag}), would be sufficient to take into account the electron density derived
from \compmag\ at the magnetic poles (see above).

The physics of shocks in these systems is however rather complicated for the intrinsic three-dimensional
nature of the problem. 
Indeed, if from one side the supersonic motion of the NS ensures the formation of a bow shock, on the
other hand this discontinuity occurs in a region where the presence of a strong magnetic field 
 plays an important role in determining the gas configuration. 

A pure dipolar magnetic field may lead to the formation of a bow shock which follows approximatively
the shape of the magnetic lines in the direction of motion of the NS, together with a channeling of  matter towards 
the magnetic poles. This accretion geometry close to the NS could be approximated with a cylindrical
accretion, suitable to be described with some accuracy by the \compmag\ model.
However, if non-negligible multipole magnetic field components are presents,  then matter
can also (or mainly) being accreted at the equator of the NS.

As shown in the data analysis, the spectra of both \src\ and \igr\ can be alternatively described
by a two-component model consisting of a soft and hard BB (see Table \ref{fits_2bb}).
The temperature of the softer BB is consistent with that of the seed photons in both \comptt\
and \compmag, while the temperature of the harder BB is comparable with the electron
temperature of the above mentioned Comptonization models.
This means that the harder BB is actually playing the role of a Comptonization feature, but the quality
of the high-energy data (see Figs. \ref{eeuf_1739} and \ref{eeuf_17544}) does not allow to distinguish
between an unsaturated (\comptt\ and \compmag) or saturated (BB) Compton regime. 

The apparent radii of both BB components are consistent with emission regions of the order
of a polar cap radius, thus pointing in favour of a very compact emission 
region. The implications for the accretion geometry in this context are that
both the regions of the soft seed photons and of the Comptonized ones
are visible.

Summarizing, the current data presently do not allow us to distinguish
whether the 0.1--60 keV \mbox{X-ray} spectra of \src\ and \igr\ form as a result of a single unsaturated Comptonization
process, or arise from two different and directly visible zones of seed and Comptonized photons.

Using the calibration of stellar parameters for Galactic O stars of \cite{Martins2005},
we find $M_{\rm SG} \sim 31~M_{\odot}$, $R_{\rm SG} \sim 21~R_{\odot}$ for \src, and 
$M_{\rm SG} \sim 30~M_{\odot}$, $R_{\rm SG} \sim 22~R_{\odot}$ for \igr, respectively.
If the orbital period of \src\ is 51.47 d, the minimum and maximum distances of the NS from the giant
companion are in the range $d_{\rm min} \sim 8-4.5 ~R_{\rm SG}$  and  $d_{\rm max} \sim 9.8-13.4 ~R_{\rm SG}$
for orbital eccentricity $e=0.1-0.5$, respectively.
These values change to $d_{\rm min} \sim 3.2-1.8 ~R_{\rm SG}$  and  $d_{\rm max} \sim 3.9-5.3~ R_{\rm SG}$
if the orbital period is  12.89 d (see Section \ref{intro}).

In the case of \igr, on the other hand, the maximum allowed eccentricity for its orbital period 
of 4.93 d is $e=0.4$, and the ranges of minimum and maximum distances to the supergiant companion 
is $d_{\rm min} \sim 1.6-1.1~ R_{\rm SG}$ and $d_{\rm max} \sim 1.9-2.4 ~R_{\rm SG}$.

The mass loss rates of the supergiant companions of \src\ and \igr\ have, within the uncertainties
with which we know the masses, radii, effective temperatures
and bolometric luminosities of the two stars,
the same value of about $2 \times 10^{-6}$\,M$_\odot$\,yr$^{-1}$ \citep{Vink2000}.

Whatever the actual orbital period of \src\ and the eccentricity of the
two systems, the distance to the supergiant companion is lower
for \igr. It is thus intriguing to observe that the optical depth derived from the Comptonization
models is higher in the latter source than in \src\ (see Tables \ref{fits_comptt} and \ref{fits_compmag}), 
as one would expect for a closer orbiting system.

\section{Conclusions}

We have analyzed  data from the outbursts of two sources, \src\ and \igr, which are
considered to be the prototypes of the SFXTs class.
During the bright flare they reached peak luminosities of  $\sim 2 \times 10^{36}$ erg s$^{-1}$
and $\sim 5 \times 10^{36}$ erg s$^{-1}$, respectively.

The presented \sw\ data do not allow us to discriminate whether the \mbox{X-ray} spectra
are the result of a single unsaturated Comptonization process, or whether we observe both regions
where seed photons are produced and Comptonization mostly takes place (two-BB model).

The electron density in the region of the \mbox{X-ray} spectral formation, 
computed to first approximation using the best-fit parameters of the \compmag\ model, is
a factor about 1000 higher than expected from the continuity equation at the 
magnetospheric radius. We propose that the formation of a bow shock at the
NS magnetosphere, due to its supersonic orbital motion around the supergiant
companion, may increase the density enough to explain the difference.
This effect needs to be investigated in a future work by means of 3D magnetohydrodynamical
simulations using the specifically devoted FLASH code.

The possible feature around 20 keV which is observed using a one-component Comptonization
model (see Fig. \ref{eeuf_17544}, bottom panel) is intriguing, as it can not be 
attributed to any particular shape of the BAT instrument effective
area, however that fact that it is strongly dumped using a two-BB model and it has a low statistical significance  
do not allow us to draw firm conclusions.

Nevertheless, this feature needs a further investigation with higher-resolution missions
in the energy range around 20 keV, in particular with NuSTAR.
If true, it would be the first measure ever obtained of the magnetic field intensity
($\sim 2 \times 10^{12}$ G) in a SFXT.

%%%%%%%%%%%%%%%%%%%%%%%%%%%%%%%%%%%%%%%%%%%%%%%%%%%%%%%%%
\section*{Acknowledgments}
%%%%%%%%%%%%%%%%%%%%%%%%%%%%%%%%%%%%%%%%%%%%%%%%%%%%%%%%%

We thank the {\it Swift} team duty scientists and science planners, and the 
remainder of the {\it Swift} XRT and BAT teams, S.\ Barthelmy 
and J.A.\ Nousek, 
in particular, for their invaluable help and support of the SFXT project as a whole. 
We thank C.\ Guidorzi for helpful discussions. 
We acknowledge financial contribution from the agreement ASI-INAF I/009/10/0
and from contract ASI-INAF I/004/11/0. 
PE acknowledges financial support from the Autonomous Region of Sardinia 
through a research grant under the program PO Sardegna FSE 2007--2013, L.R. 7/2007 
``Promoting scientific research and innovation technology in Sardinia''.
This work was supported at PSU by NASA contract NAS5-00136.

%%%%%%%%%%%%%%%%%%%%%%%%%%%%%%%%%%%%%%%%%%%%%%%%
% BIBLIOGRAPHY
%%%%%%%%%%%%%%%%%%%%%%%%%%%%%%%%%%%%%%%%%%%%%%%%

\appendix
\onecolumn
\section{The optical depth ranges of applicability of the \compmag\ model}

\noindent
The theoretical spectra of the \compmag\ model were reported by F12 for the case of optical depth $\tau <1$. 
In this section we better explain how this impacts the spectral fitting analysis of real data. 
In particular, the fits of the broad-band spectrum of \src\ with the \compmag\ model (see Table  \ref{fits_compmag}) 
yield an optical depth formally less than 1, albeit within a factor of 3 from unity. 
It is worth noting that in \compmag\ this is the vertical optical depth of
the accretion column and it is computed including an energy-independent correction to the 
Thomson cross-section, due to the presence of a magnetic field $B\ga 10^{12}$ G. 
More specifically,  it is the opacity related to  photons which diffuse across the field lines.
The relation between the optical 
depth presented in Table \ref{fits_compmag} can be written in terms of Thomson optical depth as 
$\tau\approx 10^{-3}\tau_{\rm T}$, that in the case of \src\ is $\tau_{\rm T}\sim 300$.

After rearranging some terms, equation (35) in F12 can be written as
\[
 -\frac{\sigma_\parallel}{\overline\sigma}\mathcal S(x,\tau) =-\frac{\sigma_\parallel}{\overline\sigma}\frac{v}{c}\pder{n}{\tau}+\frac{\sigma_\parallel}{\overline\sigma}\frac{1}{c}\tder{v}{\tau}\frac{x}{3}\pder{n}{x}
 +\frac{1}{3}\frac{\sigma_\parallel}{\overline\sigma}\pder{^2n}{\tau^2}-\left(\frac{\xi v}{c}\right)^2\frac{\sigma_\parallel}{\overline\sigma}n 
 +\frac{kT_e}{m_ec^2}\frac{1}{x^2}\pder{}{x}\left\{x^4\left[n+\left(1+\frac{1}{3}\frac{m_ev^2}{kT_e}\right)\pder{n}{x}\right]\right\}.
\]
In the escape time prescription provided by BW07, the spatial diffusion of photons is described by 
\be
 \frac{1}{3}\frac{\sigma_\parallel}{\overline\sigma}\pder{^2n}{\tau_\parallel^2}-\left(\frac{\xi v}{c}\right)^2\frac{\sigma_\parallel}{\overline\sigma} n=\lambda^2 n, \label{diffbw}
\ee
or, more clearly, by
\be
 \frac{1}{3}\frac{\sigma_\parallel}{\overline\sigma}\pder{^2n}{\tau_\parallel^2}-\frac{n}{r_{\rm 0}n_e\overline\sigma\tau_\bot}=\lambda^2 n, \label{diffbw_tau}
\ee
where $d\tau_\parallel\equiv d\tau=n_e\sigma_\parallel dZ$ is the optical depth along the $Z$-axis and $d\tau_\bot=n_e\sigma_\bot dr$ is the optical depth along the $r$-axis, perpendicular to the magnetic field, and $\lambda$ is the eigenvalue. The right hand side of equation \ref{diffbw_tau} can be approximated as
\be
\left(\frac{10^2}{\tau_{\rm T}}\right)^2\left[\frac{1}{3}-\frac{Z_0^2}{10^3r_{\rm 0}^2}\right]n \approx\lambda^2 n \label{diffeqncoeff}
\ee
where $d\tau_{\rm T}=n_e\sigma_{\rm T}dZ$ and since the eigenvalue $\lambda^2<1$, also the coefficient of $n$ on the right hand side of 
equation (\ref{diffeqncoeff}) should be less than 1, in order for diffusion approximation to hold. In particular, the coefficient $\mathcal D_\parallel=\sigma_\parallel/(3\overline\sigma\tau_\parallel^2)$ takes into account the spatial diffusion of photons propagating 
along the magnetic field lines, while the coefficient $\mathcal D_\bot=1/(r_{\rm 0}n_e\overline\sigma\tau_\bot)$ accounts for the diffusion of photon travelling perpendicular to the field.

Following the approach of BW07, from equation (\ref{diffeqncoeff}) it follows that the diffusion approximation for
photons propagating in the direction of the magnetic field is valid if the corresponding Thomson optical depth 
$\tau_{\rm T}\ga 10^2$. This can be actually the condition for a meaningful applicability of the \compmag\ model, 
together with the requirement that both the optical depths $\tau_\parallel$ and $\tau_\bot$ need to be larger than unity.
From equation (\ref{diffeqncoeff}) it is also evident that the ratio between the spatial diffusion in the $Z$-direction 
($\mathcal D_\parallel$) and in the $r$-direction ($\mathcal D_\bot$) depends upon the choice of the column size ($Z_0/r_{\rm 0}$).

\bsp
\label{lastpage}
\end{document}